\documentclass[letters,fleqn,usenatbib,useAMS]{mnras}

\usepackage{graphicx}	
\usepackage{amsmath}	

\title[Pulsation variability of $\gamma$\,Equ]{
The rapidly oscillating Ap star $\gamma$\,Equ: linear polarization as an enhanced pulsation diagnostic? 
 }

\stepcounter{footnote}

\author[Hubrig et al.]{
S.~Hubrig$^{1}$\thanks{Corresponding author: shubrig@aip.de},
S.~P.~J\"arvinen$^{1}$
I.~Ilyin$^{1}$,
K.~G.~Strassmeier$^{1}$,
M.~Sch\"oller$^{2}$
\\
$^{1}${Leibniz-Institut f\"ur Astrophysik Potsdam (AIP), An der Sternwarte~16, 14482~Potsdam, Germany} \\
$^{2}${European Southern Observatory, Karl-Schwarzschild-Str.~2, 85748 Garching, Germany} \\
}

\date{Accepted XXX. Received YYY; in original form ZZZ}

\pubyear{2021}

\begin{document}
\label{firstpage}
\pagerange{\pageref{firstpage}--\pageref{lastpage}}
\maketitle

\begin{abstract}
We present the first short time scale
observations of the roAp star $\gamma$\,Equ in linear polarized light 
obtained with the PEPSI polarimeter installed at the LBT.
These observations are used to search 
for pulsation variability in Stokes~$Q$ and $U$ line profiles belonging to different elements.
The atmospheres of roAp stars are significantly stratified  
with spectral lines of different elements probing different atmospheric depths.
roAp stars with strong magnetic fields, such as $\gamma$\,Equ with
a magnetic field modulus of 4\,kG and
a pulsation period of 12.21\,min,  are 
of special interest because the effect of the magnetic field on the structure 
of their atmospheres can be studied with greatest detail and accuracy. 
Our results show that we may detect changes
in the transversal field component in \ion{Fe}{i} and rare-earth lines possessing large 
second-order Land\'e factors. 
Such variability can be due to the impact of pulsation on the transverse magnetic field,
causing changes in the obliquity angles of the magnetic force lines.
Further studies of roAp stars in linear polarized light and subsequent detailed modelling
are necessary to improve our understanding of the involved physics.
\end{abstract}

\begin{keywords}
  techniques: polarimetric --- 
   techniques: spectroscopic --- 
  stars: individual: $\gamma$\,Equ ---
  stars: magnetic fields ---
  stars: oscillations (including pulsations)
\end{keywords}



\section{Introduction}
\label{sec:intro}

A number of cool Ap stars belong to a subgroup called rapidly oscillating Ap (roAp) stars.
The first discovery of a pulsating Ap star was reported by \citet{Kurtz1978}, who detected a 12\,min  pulsation 
period in Przybylski's star (HD\,101065) using ground-based photometric observations.
roAp stars pulsate in  high-overtone, low-degree, nonradial p-modes modes with periods in 
the range of about 5 to 24\,min.
Compared to the several thousand 
known Ap stars (e.g.\ \citealt{RensonManfroid2009}), roAp stars are relatively rare with less than hundred known to
date \citep[e.g.][]{Smalley2015,Holdsworth2021}.

Apart from photometric variability, roAp stars show rapid radial velocity variations with periods similar
to those obtained from photometry. 
The highest pulsation amplitudes of radial velocities were detected for lines of rare earth 
elements and the H$\alpha$ core (e.g.\ \citealt{Kurtz2003}, and references therein), indicating that
the pulsation amplitudes are a function of atmospheric depth. While no noticeable radial velocity variations
are detected in lines belonging to iron, which is usually concentrated in deeper atmospheric layers,
the pulsations observed in radial velocity variations of lines of rare 
earth elements arise at atmospheric heights where the magnetic pressure exceeds the gas pressure and the Alfv\'en 
velocity is greater than the acoustic velocity. This means that the observed pulsations are primarily magnetic with an
acoustic component. Spectroscopic observations of the variability of spectral lines belonging to 
different elements formed in the outer atmospheric
layers of these stars offer unique possibilities for investigating details of the physics of 
propagating magnetoacoustic waves. Furthermore, using different spectral lines it is possible to resolve the 
pulsation behaviour as a function of optical depth over a large range with log\,$\tau_{\rm 5000}$ from $-$5 to 0
(e.g.\ \citealt{Mathys2007}).

According to numerous theoretical and observational studies, roAp stars are oblique pulsators and
the strength and geometry of their global,
approximately dipolar magnetic fields constrain the pulsation modes.
Their non-radial pulsations have
axes that are inclined to the rotation axis of the star and nearly aligned
with  the  magnetic axis \citep{Kurtz1982,ShibahashiTakata1993,SaioGautschy2004,BigotKurtz2011}.
As the star rotates,
a different pulsation aspect is seen along the line of sight,
leading to observed amplitude and phase modulations. These
modulations together with the light curve can provide information on the geometry of the observed pulsations.
The pulsation amplitude and phase changes have recently been modeled by \citet{Quitral2018}, who showed them
as a function of atmospheric depth and discussed how they
behave between the pole and equator as a consequence of the
interaction of the acoustic and magnetic components.
In their work, the magnetic component was assumed to be a wave that dissipates inside the star.

Assuming that the pulsations of roAp stars are to a large extent governed
by their magnetic field, \citet{Hubrig2004} for the first time measured the magnetic field variation 
over the pulsation cycle in six roAp stars using the low-resolution FORS\,1 (FOcal Reducer low dispersion 
Spectrograph; \citealt{Appenzeller1998}) instrument
mounted on the 8-m Melipal (UT3) telescope of the VLT to study how the magnetic field and 
the pulsations interact. Only the roAp star HD\,101065, 
which has one of the highest photometric pulsation amplitudes, showed a signal of magnetic variability with a 
frequency of 1.365\,mHz and an amplitude of $39\pm12$\,G. First theoretical considerations by 
\citet{Hubrig2004} demonstrated
that in the case of HD\,101065 the pulsation could give rise to magnetic field variations
 $\left|\frac{\delta B_\theta}{B}\right|\sim0.1 \rightarrow \left|\delta B_\theta\right|\approx 100$\,G, 
consistent with a possible detection. Unfortunately, the second attempt to measure magnetic 
variability in this star over 4 hours failed to obtain a positive detection, due to the higher 
noise level of the new observations.

To enhance the diagnostics of pulsations in roAp stars, \citet{Mathys2005,Mathys2007} studied the pulsation 
behaviour of the individual $\pi$ and $\sigma$ components using 
the \ion{Eu}{ii}\,$\lambda$6437 line in a spectroscopic time series of the strongly magnetic roAp star HD\,166473
exhibiting radial velocity variations due
to pulsation with three frequencies, 1.833, 1.886, and 1.928\,mHz.  As this star has a very long rotation 
period of about 10.5\,yr \citep{Mathys2020}, and possesses a strong surface magnetic field of
about 9\,kG, several spectral lines are magnetically resolved with large wavelength separations of 
the Zeeman split components. 
Due to HD\,166473's very anomalous atmospheric structure it became possible to study for the first time the vertical
resolution of the pulsation modes into standing waves in the
atmosphere and overlying running waves in the upper atmosphere \citep{Kurtz2003}. 
The results of \citet{Mathys2005,Mathys2007} 
hinted at the occurrence of variations of
the mean magnetic field modulus with a pulsation frequency of 1.928\,mHz and an amplitude of $21\pm5$\,G.

Linear polarization measurements have not been considered so far
as an additional diagnostic to advance asteroseismic modelling of pulsations in magnetic roAp stars
by studying the variability of the transverse component of the magnetic field over the pulsation cycle.
The presence of periodical variable linear polarization in nonradially pulsating hot stars, such as $\beta$~Cep
stars with significant photospheric electron-scattering opacity, was previously discussed by \citet{Odell1979}
and \citet{Stamford1980}.
The detection of linear polarization in the $\beta$~Cep star BW\,Vul with large radial
velocity and photometric pulsation amplitudes was reported by \citet{Odell1981}.
The distortion of this star is caused by nonradial pulsations leading to temporal polarization variability.
However, as for roAp stars, due to their rather low photometric and radial velocity pulsation amplitudes, 
usually of the order of mmag and 
a few hundred and less m\,s$^{-1}$, we do not expect any significant deviation from
spherical symmetry and can reasonably assume that any observed linear polarization
measured in individual spectral lines can be ascribed to the Zeeman effect. 

One very bright roAp star, the A9VpSrCrEu star $\gamma$\,Equ (=HD\,201601)
with $m_{\rm V}=4.7$, exhibiting very sharp 
lines due to the extremely long rotation period of at least 95.5\,yr \citep{Savanov2018,Bychkov2016} 
and possessing a rather strong very slowly varying surface 
magnetic field of up to 4\,kG (e.g. \citealt{Mathys2017}), is frequently observed in all four Stokes parameters 
with different spectropolarimeters to investigate the presence of cross-talk 
between linear polarization and circular polarization and vice versa.
$\gamma$\,Equ exhibits a rather high amplitude of pulsation line 
profile changes exceeding 1000\,m\,s$^{-1}$ in individual REE spectral
lines and is known to have multiple pulsation periods near 12\,min
\citep{Kurtz1983,Martinez1996}. 
\citet{Gruberbauer2008} identified seven frequencies 
that  they  associated  with five high-overtone  p-modes  and 1st and 2nd harmonics of the
dominant p-mode. Two closely spaced frequencies with the highest amplitudes correspond to the periods 
12.214 and 12.206\,min. 

Cross-talk monitoring using observations of $\gamma$\,Equ
has been performed in recent years for the two dual-beam Stokes~$IQUV$
polarimeters of the Potsdam Echelle Polarimetric and Spectroscopic Instrument (PEPSI;
\citealt{Strassmeier2015}) installed at 
the $2\times8.4$\,m Large Binocular Telescope (LBT). Due to the large aperture
of the LBT, a few linear polarization
observations of $\gamma$\,Equ were obtained with exposure times accounting for a fraction of the pulsation cycle 
of 12.2\,min, allowing us for the first time to investigate the variability of the Stokes~$Q$ and $U$ parameters 
on a short time scale of the order of minutes. 
In the following, we present the available PEPSI linear polarization observations of $\gamma$\,Equ and discuss
their potential usefulness 
for the improvement of our understanding of the pulsation properties of roAp stars.

\section{Observations and results}
\label{sect:obs}

Both polarimeters installed at the LBT are based on a classical dual-beam design
with a modified Foster prism as linear polarizer with two orthogonally polarized beams 
(ordinary and extraordinary) exiting in parallel.
During spectropolarimetric  observations with PEPSI, two pairs of octagonal 200\,$\mu$m
core-diameter fibers per polarimeter
feed the ordinary and extraordinary polarized beams per telescope via a five-slice image 
slicer per fiber into the spectrograph.
For circular polarization measurements, a quarter-wave retarder is inserted
into the optical beam in front of the Foster prism. This design minimizes the cross-talk
between circular and linear polarization to less than 0.1\,per cent \citep{Ilyin2012}. The study of the instrumental
polarization determined on the basis of unpolarized and polarized standard
stars resulted in a very low value, $10^{-5}$ (Ilyin, priv.\ comm.).
The low level of cross-talk and 
instrumental polarization demonstrates that PEPSI has the potential to reliably recover Zeeman signatures.
Three grisms per spectrograph arm cover the wavelength range from 3837 to 9067\,\AA{}.
All polarimetric spectra have a fixed spectral resolution of $R\sim130\,000$ corresponding to 0.06\,\AA{}
at $\lambda$7600\,\AA{}. For data reduction, the software package SDS4PEPSI
(Spectroscopic Data Systems for PEPSI) based on \citet{Ilyin2000} is used.

\begin{figure}
\centering 
\includegraphics[width=0.375\textwidth]{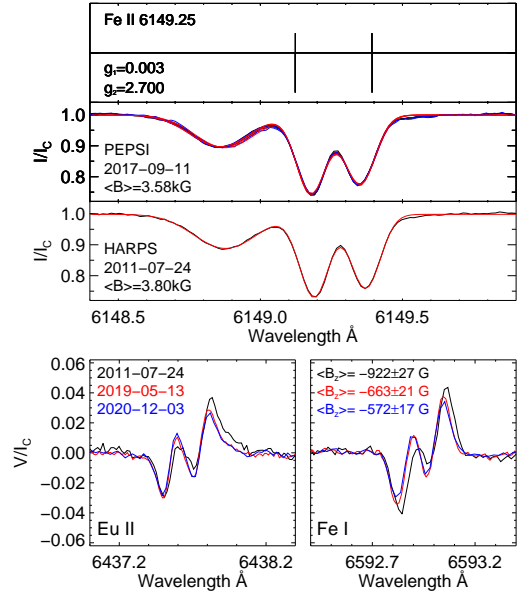}
        \caption{
{\it Top:} The Zeeman doublet \ion{Fe}{ii}~$\lambda$6149.25 used for the measurements of the mean magnetic field modulus.
The solid blue and black lines present the observed line profiles whereas the fits obtained using multiple Gaussians are 
represented by red lines.
{\it Bottom:} Significant changes observed in the 
Stokes~$V$ profiles of the lines \ion{Eu}{ii}~$\lambda$6437.63 and \ion{Fe}{i}~$\lambda$6592.91 recorded
in different years.
}
   \label{fig:field}
\end{figure}

According to \citet{Mathys2017} and \citet{Savanov2018},
in recent years, both the mean longitudinal magnetic field
and the magnetic field modulus of $\gamma$\,Equ are gradually decreasing in absolute value.
Our measurements of the field modulus and 
longitudinal field using PEPSI observations from 2017, 2019, and 2020 and one ESO archival observation using
HARPS\-pol obtained on 2011 July 24 confirm their results. In Fig.~\ref{fig:field} 
we present on the bottom significant changes in the 
Stokes~$V$ profiles for the lines \ion{Eu}{ii}~$\lambda$6437.63 and \ion{Fe}{i}~$\lambda$6592.91. 
No Stokes~$V$ spectra were obtained in this region in 2017.
To measure the mean 
longitudinal field, we applied the least-squares deconvolution (LSD) technique \citep{Donati1997} using a 
line mask containing six blend-free \ion{Fe}{i} lines. 
Magnetically split components of the Zeeman doublet \ion{Fe}{ii}~$\lambda$6149.25 line used for the measurements of 
the mean magnetic field modulus in the years 2011 and 2017 are presented on the top 
of Fig.~\ref{fig:field}. No PEPSI observations at this wavelength were carried out in 2019 and 2020. 

Two Stokes~$Q$ spectra
over one pulsation cycle and two Stokes~$U$ spectra
over two consecutive pulsation cycles were recorded on 2017 September 11, when the LBT was used in binocular mode
using both mirrors referred as SX and DX and crossdispersers~3 and 5 to cover the wavelength regions 
4800--5441\,\AA{} and 6278--7419\,\AA{}.
As the Zeeman effect strongly depends on the wavelength,
we focused in this study on the observations obtained at the longer wavelengths in the range 6278--7419\,\AA{}.
The signal-to-noise ratio ($S/N$) in the corresponding Stokes~$I$ spectra was 500--700.
Conventionally, the Stokes~$V$ exposures in circular polarized light are obtained for retarder angles 
$45^{\circ}$ and $135^{\circ}$, whereas Foster prism 
position angles $0^{\circ}$ and $90^{\circ}$ are used to obtain Stokes~$Q$ spectra and Foster prism 
position angles $45^{\circ}$ and $135^{\circ}$ for Stokes~$U$ spectra with respect to the north-south direction.
Given the low cross-talk measured for the PEPSI polarimeters, only respectively one position angle
was used to record Stokes~$Q$ and Stokes~$U$ spectra in observations carried out on 2017 September 11. 
Accordingly, no signal was detected in Stokes~$Q$ and $U$ profiles of numerous telluric lines around 6880\,\AA{}
and in the Zeeman doublet \ion{Fe}{ii}\,$\lambda$6149.258 recorded during the same night in the wavelength region
5441--6127\,\AA{} using crossdisperser~4. 

The first observation of the Stokes~$Q$ spectrum recorded with the DX mirror started on 2017 September 11 at 
UT05:38:53.6 with an exposure time of 2\,min and the second observation at UT05:47.04.1 with an exposure time of
6\,min.
Using the SX mirror, the first observation of the Stokes~$U$ spectrum started at 
UT05:42:34.2 with an exposure time of 2\,min
and the second observation at UT05:54:43.8 with an exposure time of 6\,min.
Given these times of observations and the pulsation period of about 12.21\,min,
the second Stokes~$Q$ spectrum has a phase shift of 0.83 with respect to the first Stokes~$Q$ exposure
and the second Stokes~$U$ spectrum a phase shift of 1.16 with respect to the first Stokes~$U$ exposure.
All phases are calculated for the middle of the exposures.

\begin{figure}
\centering 
\includegraphics[width=0.235\textwidth]{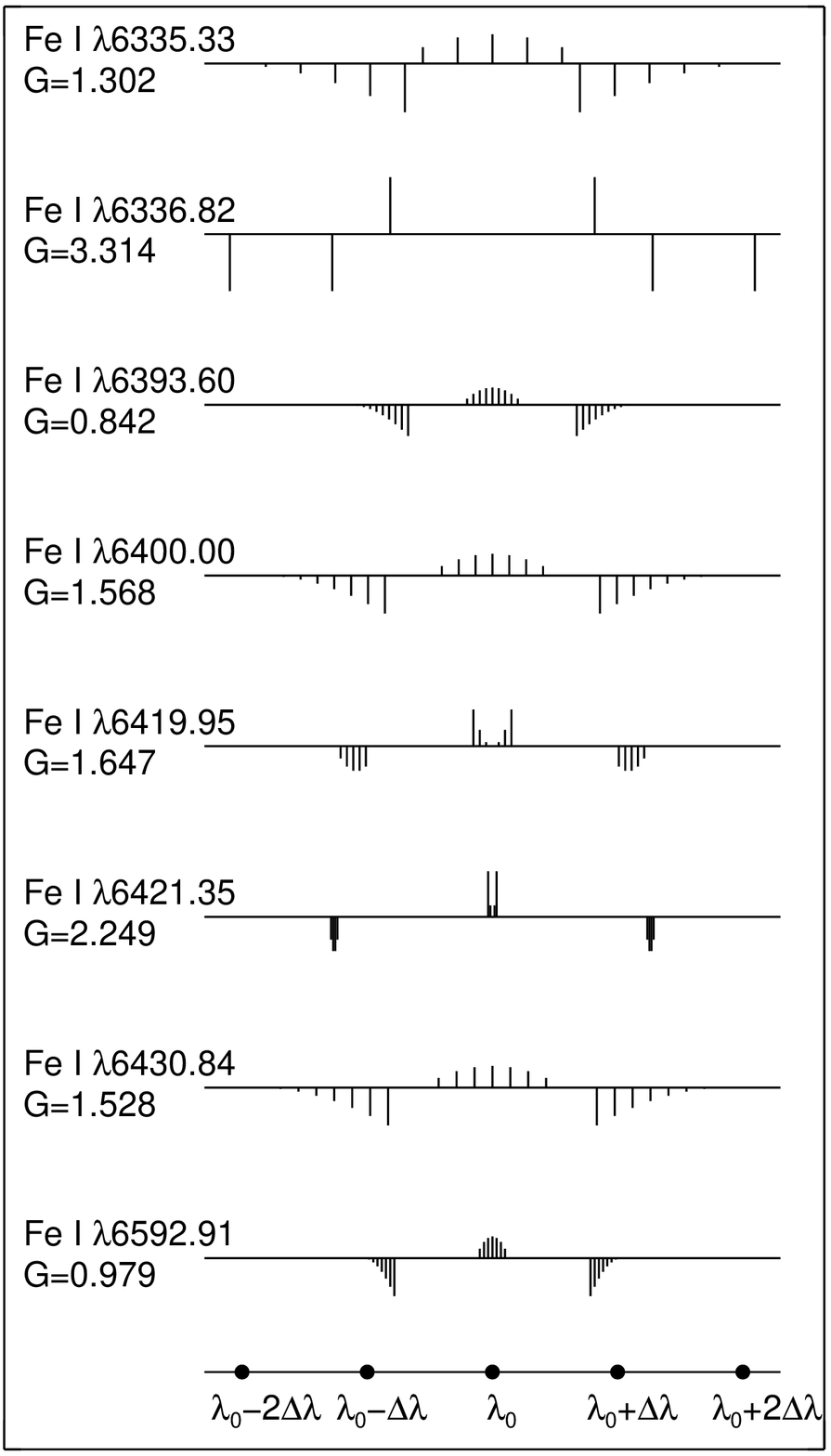}
\includegraphics[width=0.235\textwidth]{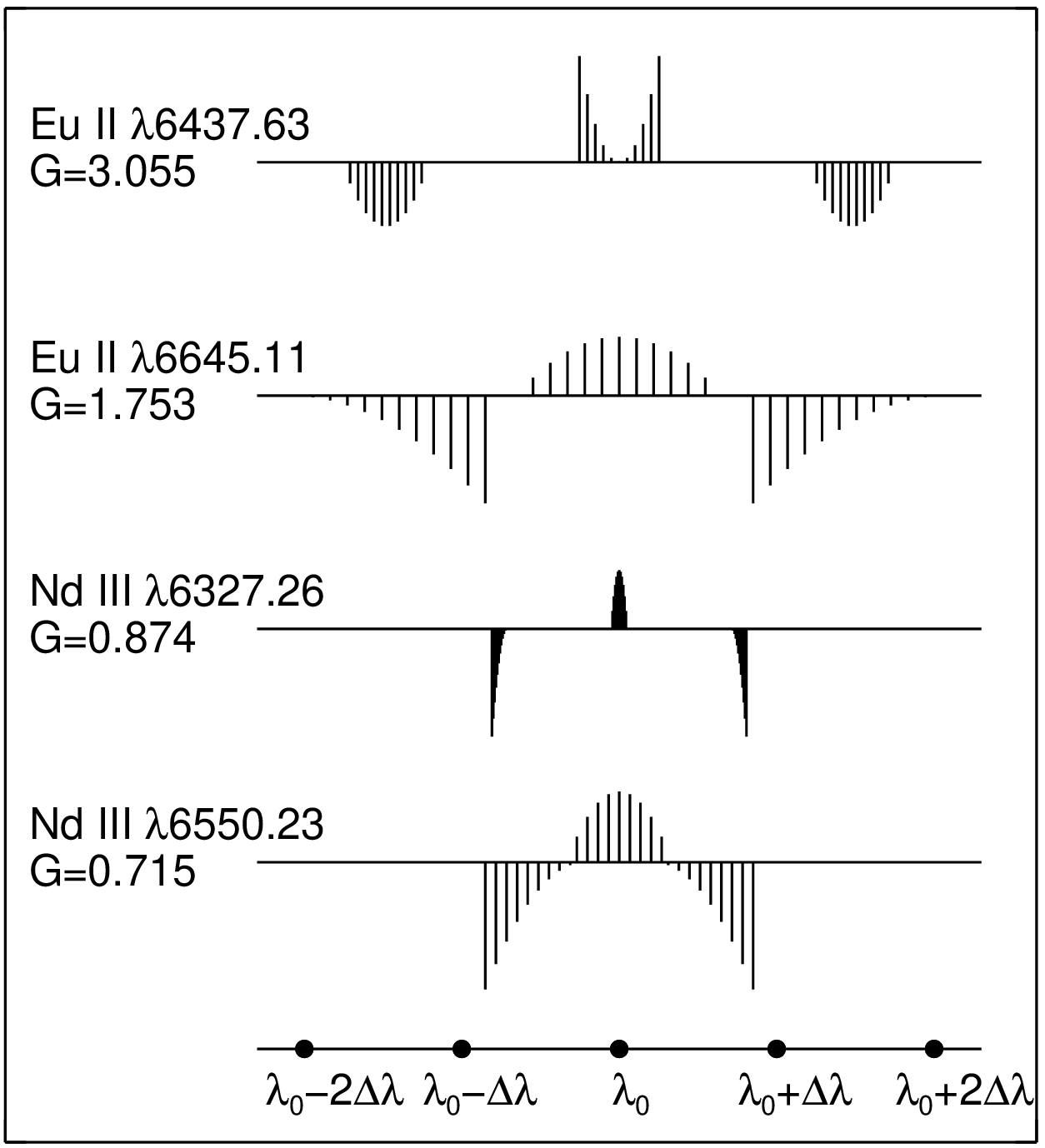}
        \caption{
Zeeman patterns of six blend-free \ion{Fe}{i} lines (left panel) and of four lines belonging to rare earth elements
(right panel) with the corresponding second-order effective Land\'e factors G.
Each Zeeman component is represented by a vertical bar whose length is proportional to its relative strength.
The $\pi$ components are presented above the horizontal wavelength axis,
the $\sigma_{\pm}$ components below it.
The unit length of the wavelength axis is $\Delta\lambda_B$ defined as being the wavelength
shift from the line centre of the $\sigma$ components in a normal Zeeman triplet with a Land\'e factor $g=1$.
}
\label{fig:fepat}
\end{figure}

\begin{figure*}
 \centering 
\includegraphics[width=0.700\textwidth]{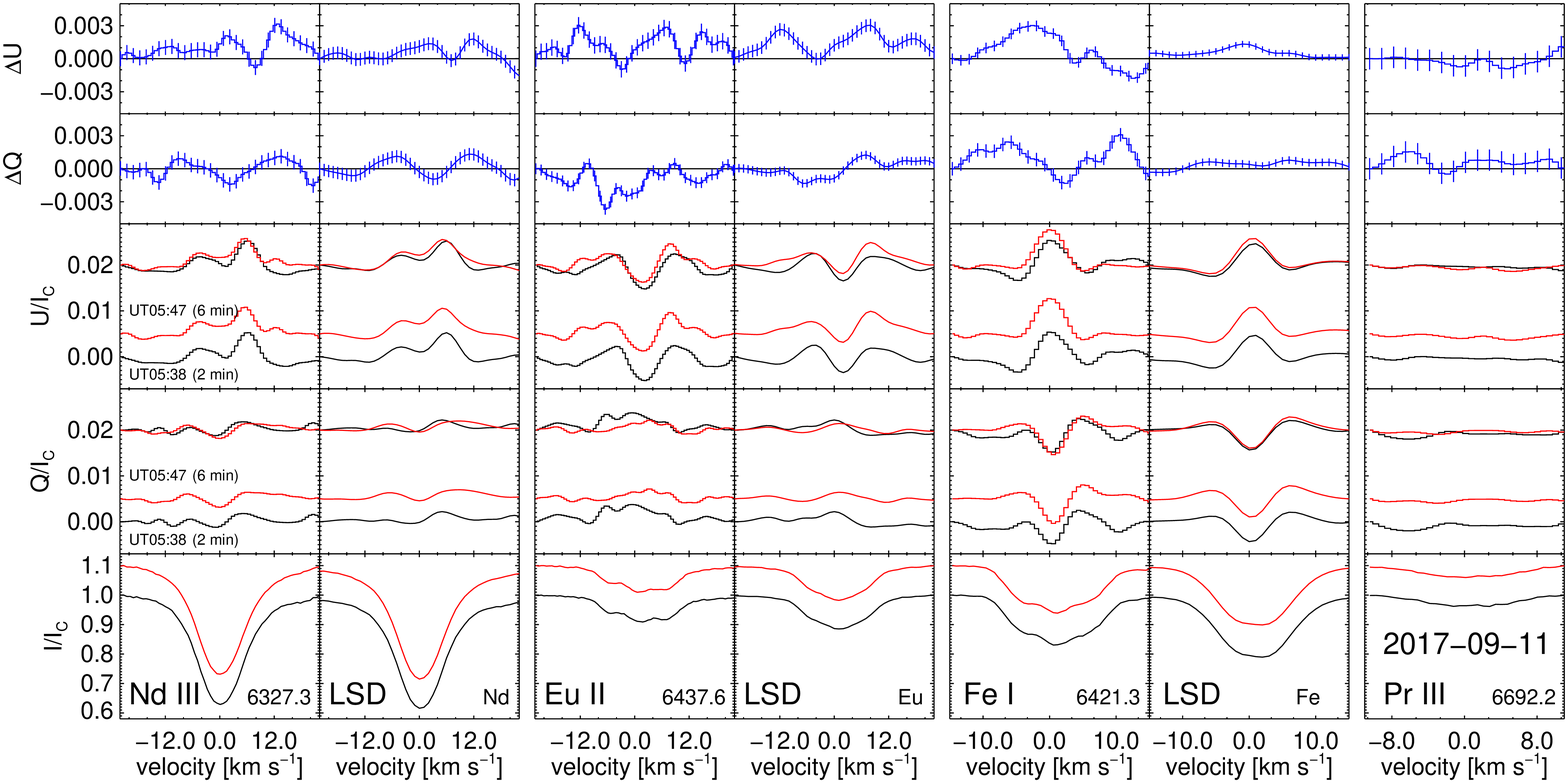}
        \caption{
Zeeman signatures of $\gamma$\,Equ in the linear polarization line profiles of different lines recorded
with PEPSI on two different pulsation phases in 2017 September 11.
Individual and overplotted Stokes~$I$ profiles for single and LSD profiles are shown in the bottom panels followed by 
individual and overplotted Stokes~$Q$ and $U$ profiles in the middle panels.
The upper panels present the differences between the Stokes~$Q$ and $U$ profiles
with the associated error bars.
Since the spectral resolution of $R\sim130\,000$ offered by the PEPSI observations
is sampled by 4.2 CCD pixels, 
to achieve a higher $S/N$, the Stokes~$Q$ and $U$ spectra have been smoothed using Gaussians. 
}
   \label{fig:lin}
\end{figure*}

\begin{figure}
 \centering 
\includegraphics[width=0.430\textwidth]{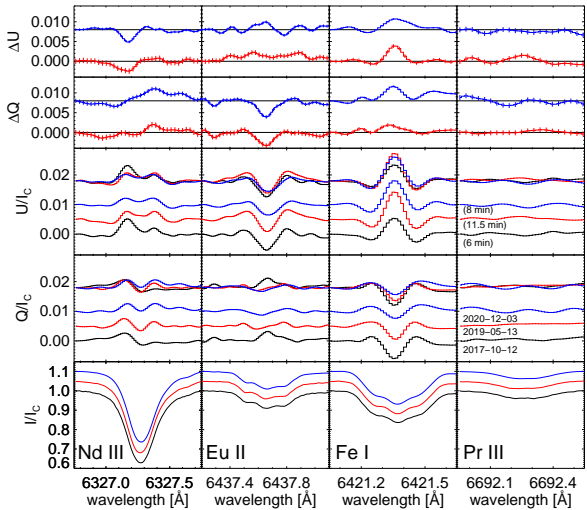}         
\caption{
 Same as in Fig.~\ref{fig:lin}, but for individual linear polarization observations of $\gamma$\,Equ obtained 
 on 2017 October 12, 2019 May 13, and 2020 December 3
 with exposure times of 6, 11.5, and 8\,min, respectively. The differences in the Stokes~$Q$ and $U$ profiles
 presented in the upper panels are calculated relative to the
observations obtained on 2017 October 12.
}
   \label{fig:lin1}
\end{figure}

As Stokes~$Q$ and $U$ are
related to the second derivative of Stokes~$I$ \citep{Landi2004}, their signatures in the line profiles 
are usually much smaller than the associated Zeeman signatures in circular polarization.
Therefore, to detect their pulsation variability, we searched in the observed spectra
for diagnostic blend-free spectral lines with large second-order effective Land\'e 
factors G. According to 
\citet{Landi2004}
$G = g_{\rm eff}^2 - \frac{1}{80}\left( g_1 - g_2\right)^2(16s-7r^2-4)$ with 
$g_{\rm eff} = \frac{1}{2}\left( g_1 + g_2\right) + \frac{1}{4} \left( g_1 - g_2\right) r$,
$r = J_1\left(J_1+1 \right) - J_2\left(J_2+1 \right)$, and
$s = J_1\left(J_1+1 \right) + J_2\left(J_2+1 \right)$,
with $g_1$ and $g_2$ the Land\'e factors
and $J_1$ and $J_2$ the total angular momentum quantum numbers, of the upper and lower levels, respectively.
Examples of Zeeman patterns of eight selected \ion{Fe}{i} lines and of four lines belonging to rare earth
elements are presented in Fig.~\ref{fig:fepat}, together with the corresponding second-order 
effective Land\'e factors G. Among the \ion{Fe}{i} lines, the strongest sensitivity to linear polarization
show the pseudo-doublet \ion{Fe}{i}~$\lambda$6336.82 and pseudo-triplet \ion{Fe}{i}~$\lambda$6421.35 with G factors of
3.314 and 2.249, respectively. 
Among the two most clean blend-free Eu lines, the pseudo-triplet \ion{Eu}{ii}~$\lambda$6437.63 has the strongest G value of 3.055
and, among the blend-free Nd lines, the pseudo-triplet \ion{Nd}{iii}~$\lambda$6327.26 has a G value of 0.874.

Although the line profiles in Stokes~$Q$ and $U$ spectra were not recorded at the same pulsation phases,  
the obtained observations offer us an excellent opportunity for a differential comparison between 
consecutive Stokes~$Q$ and $U$ spectra separately.
In Fig.~\ref{fig:lin} we display Stokes~$Q$ and $U$ profiles of individual lines and the LSD 
profiles for the Fe, Nd and Eu lines presented in Fig.~\ref{fig:fepat}. As a significant fraction of the \ion{Fe}{i} lines
exhibits a pseudo-triplet Zeeman structure, we show in this figure the pseudo-triplet
\ion{Fe}{i}~$\lambda$6421.35 line.
In the last panel on the right side we display the 
Stokes~$Q$ and $U$ profiles for the \ion{Pr}{iii}~$\lambda$6692.25 line with a low second-order effective Land\'e 
factor G of 0.19 chosen as a control null line. The upper panels in Fig.~\ref{fig:lin} show the
difference between 
the recorded Stokes~$Q$ and $U$ profiles on two different pulsation phases with the associated error bars.
Several Stokes~$Q$ and $U$ profiles exhibit a clear variability above the noise level, whereas no
significant variability is detected for the \ion{Pr}{iii}~$\lambda$6692.25 line.
Notably, the LSD profiles for
Nd and Eu constructed using line masks with two Nd and two Eu lines, respectively,  show the same variability
character as the profiles
for single Nd and Eu lines. Obviously, future observations should be carried out in a more extended
spectral region to permit the selection of a larger number of lines with similar
Zeeman structure and large second-order effective Land\'e factors for the analysis.

Three additional individual linear polarization observations of $\gamma$\,Equ were obtained on
2017 October 12, 2019 May 13, and 2020 December 3 using two retarder angles for both Stokes~$Q$ and $U$,
with exposure times of 6, 11.5, and 8\,min, respectively, with a $S/N$ in the corresponding Stokes~$I$
spectra of about 1000. The differences in Stokes~$Q$ and $U$ profiles are calculated relative to the
observations obtained on 2017 October 12.
Observations presented in Fig.~\ref{fig:lin1}, display a much stronger variability of both the Stokes~$Q$ 
and $U$ profiles of studied Nd, Eu, and Fe individual lines. While the Stokes~$Q$ and $U$ observations in 2019 were obtained 
with an exposure time corresponding to the coverage of almost the 
full pulsation cycle, the observations on 2017 October 12 have a coverage of about half of the cycle and those
on 2020 December 3 of about two thirds of the cycle.
Assuming phase 0 for the middle of the first $Q$ and $U$ 
observations on 2017 September 11 and a pulsation period  
corresponding to the frequency $f_{1}=1.364594$\,mHz, which has the highest amplitude \citep{Gruberbauer2008}, 
the observations on 2017 October 12 started at 
UT03:13:03.5 for Stokes~$Q$ and those at UT03:20:42.3 for Stokes~$U$ would correspond 
to the pulsation phases 0.153 and 0.477 for Stokes~$Q$ and $U$, respectively. 
The phases for the Stokes~$Q$ and $U$ observations obtained on 2020 December 3 
are difficult to estimate due to the long time lapse since 2017.
The comparison 
of the Stokes~$Q$ and $U$ profiles presented in Figs.~\ref{fig:lin} and \ref{fig:lin1} shows a distinct variety
of profile shapes, suggesting that the observations acquired in 2017 and 2020
correspond to different pulsation phases. We note that due to the very long rotation period of $\gamma$\,Equ
of almost a century, the possibility that the observed changes in the  Stokes~$Q$ and $U$ profiles
are caused by the changing aspect of the overall geometry of the star during the rotation cycle can be disregarded.

\section{Discussion}
\label{sec:meas}

While the Stokes~$V$ parameter depends upon the magnetic vector projection onto the line of sight,
the Stokes~$Q$ and $U$ parameters depend upon the magnetic vector projected onto the plane perpendicular to the 
line of sight. Previous searches for Stokes~$V$ variability over the pulsation cycle for $\gamma$\,Equ
were inconclusive. \citet{Leone2003}  reported a clear detection of mean longitudinal magnetic field variability of
up to $240\pm37$\,G in the \ion{Nd}{iii}~$\lambda$5845.07 line over the pulsation cycle. However,
this discovery was questioned by \citet{Kochukhov2004}, who obtained an upper limit of 40--60\,G using
13 \ion{Nd}{iii} lines. 
Our results show that it is possible that
we detect changes in the transversal field component in Fe, Nd, and Eu lines. 
Such variability can be due to the impact of pulsations on the transverse magnetic field, causing changes in the 
obliquity angles of magnetic force lines. Since Fe is reported
to be settled in the atmospheres of roAp stars and the rare-earth elements are usually 
concentrated in the upper atmospheric layers, it is expected that the lines of different elements 
trace the pulsation amplitudes in the horizontal direction differently.
According to \citet{Saio2005}, the horizontal component of the pulsation 
velocity can be comparable to the vertical component over part of the surface. 
The possible presence of different pulsation amplitudes of the $\pi$ and $\sigma$ components 
sampling the various parts of the stellar surface differently can also to a certain degree affect the 
polarised line profiles \citep{Mathys2007}.
Pulsation-phase-resolved spectropolarimetric 
observations will permit to untangle the contribution of this effect from the contribution of variations of 
the magnetic field with the pulsation period.
Further studies of roAp stars in linear polarized light with much better time resolution of the pulsation cycle 
and subsequent detailed modelling are necessary to improve our understanding of the involved physics.

roAp stars with strong magnetic fields are of special interest because the effect of the magnetic field on 
the structure 
of their atmospheres can be studied with greatest detail and accuracy. 
Over the last decades roAp stars became extremely promising targets for asteroseismology, which is a most powerful tool
for testing theories of stellar structure. 
Admittably, the atmospheric structure of roAp stars is rather complex.
The observed element stratification affects
the radiative transfer in the atmosphere, in which, as a result,
the temperature gradient and the flux redistribution become non-standard. 
An indication of the existence of a relation between the magnetic field strength and its orientation 
and vertical element stratification was recently presented 
by \citet{Jarvinen2020} in their study of the strongly magnetic Ap star HD\,166473. Furthermore, a number of 
studies indicate the possible presence of  vertical magnetic 
field gradients (e.g. \citealt{Nesvacil2004,Kudryavtsev2012,Hubrig2018}). 
Doubtlessly, the peculiar atmospheres of magnetic roAp stars offer a unique possibility to build a
complete 3D model of a pulsating stellar atmosphere by measuring all four Stokes parameters of the spectral lines
belonging to different elements located at various heights in the atmosphere.

\section*{Acknowledgements}
We thank Gautier Mathys, Don Kurtz, and the anonymous referee for their valuable comments.
PEPSI was made possible by funding through the State
of Brandenburg (MWFK) and the German Federal Ministry of Education and
Research  (BMBF)  through  their  Verbundforschung  grants  05AL2BA1/3  and
05A08BAC.
LBT Corporation partners are
the University of Arizona on behalf of the Arizona university system;
Istituto Nazionale di  Astrofisica,  Italy;
 LBT  Beteiligungsgesellschaft,  Germany,  representing  the
Max-Planck Society, the Leibniz-Institute for Astrophysics Potsdam (AIP), and
Heidelberg  University; 
the  Ohio  State  University;
 and  the  Research  Corporation,
on behalf of the University of Notre Dame, the University of Minnesota and
the University of Virginia. 
SPJ is supported by the German Leibniz-Gemeinschaft, project number P67-2018.

\section*{Data Availability}

The PEPSI data can be obtained from the authors upon request.

\bsp	
\label{lastpage}
\end{document}